# Electronic Structure Calculations Using the Thomas-Fermi Model

Gregory C. Dente

*GCD Associates, 2100 Alvarado NE, Albuquerque, NM 87110*

Michael L. Tilton

*Boeing LTS, P.O. Box 5670, Kirtland AFB, NM 87185*

**Abstract**

Using the Wentzel-Kramers-Brillouin method, we derive a modified form of the Thomas-Fermi approximation to electron density.  This new result enables us to calculate the details of the self-consistent ion cores, as well as the ionization potentials for the first s-orbital bound to the closed-shell ion core of  the Group III, IV and V elements.  Next, we demonstrate a method for separating core electron densities from valence electron densities.  When we calculate the valence kinetic energy density, we show that it separates into two terms:  the first exactly cancels the potential energy of the ion core in the core region; the second represents the residual kinetic energy density resulting from the valence density alone.  Furthermore, we show that the kinetic cancellation and the residual kinetic energy can be derived from a slowly varying envelope approximation for the valence orbitals in the core region.  This kinetic cancellation in the core region and the residual valence kinetic energy term allow us to write a functional for the total valence energy dependant only on a low spatial frequency valence density.  In the limit, when we can freeze the potential of the closed-shell ion cores, assuming that they are not greatly influenced by the readjustment  of the valence electrons, we can minimize the total valence energy with respect to the valence density degrees of freedom.  The variation of the valence total energy equation provides the starting point for a large number of atomic, molecular and solid-state electronic structure problems.  Here, we use it to calculate the band structures resulting from the self-consistent valence density and potential on the zinc-blende and diamond lattices. We give band structure results for most of the Group III-V, as well as Group IV, materials.

## I. Introduction

Calculating the properties of atoms, molecules and solids has been one of the primary objectives of physics for the last century. Certainly, by the early 1930's, the calculating machinery of quantum mechanics was well-understood and spectacularly successful when applied to one- and two- electron systems. However, as researchers began to tackle the N-electron problems, the many-electron theories quickly became complicated and unwieldy. One powerful approach, variational calculations based on determinant wave functions, led to a set of N coupled integral-differential equations for N single-electron orbitals. These nonlinear Hartree-Fock equations then had to be solved self-consistently. Although many results have been obtained over the last eighty years, when the number of electrons became large, these procedures proved difficult. In fact, for large N, one could question the very practicality of an antisymmetric N-electron wavefunction that is a function of 3N independent variables. [1]

As an alternative to solving for an N-electron wavefunction, researchers also developed methods that dealt directly with the electron density. These density-functional theories, DFT, can be derived from, or at least approximated from, the N-electron wave equation. [1] The earliest example of a DFT was developed in the late 1920's; this is the Thomas-Fermi model, one of the earliest schemes for solving the N-electron problem while enforcing the Pauli exclusion principle and wave-particle duality. In this model, the local electron density is related to the Fermi momentum as

$$\rho(\vec{r}) = 2 \cdot \frac{4\pi P_F^3(\vec{r})}{3(2\pi\hbar)^3} = \frac{(2m(F-V(\vec{r})))^{3/2}}{3\pi^2 \hbar^3} \quad , \tag{1}$$

in which the Fermi momentum, $P_F$, of the most energetic electrons is specified by the Fermi energy, F, and the local potential, V. When the Poisson potential, as well as the exchange/correlation components of the potential, could be determined by the density, a self-consistent solution was then possible.[3] Unfortunately, Thomas-Fermi has always been considered a crude approximation that was not accurate enough for quantitative chemistry or material science calculations.[1] This paper will suggest that a slightly modified version of Thomas-Fermi can be a practical model that consistently yields high accuracy. Additionally, the modified Thomas-Fermi method leads to two remarkably helpful developments. First, we can easily separate valence electron densities from core electron densities. Second, we can show that the valence kinetic energy density can be separated into a term that exactly cancels the core potential in the core region, while the remaining term can be interpreted as a residual kinetic energy density generated by the slowly varying envelopes of the valence orbitals. This type of kinetic cancellation, based on a slowly varying envelope approximation for the valence orbitals, differs from the standard approach that emphasizes orthogonality of the valence and core electron orbitals.[2]

In Section II, supplemented by Appendix A of this paper, we will use the Wentzel-Kramers-Brillouin method, WKB, to derive an improved form of the Thomas-Fermi density that corrects the density of electrons near the atomic nucleus. In the following Section III, we will present an application which allows us to calculate a set of ionization potentials for the ionized Group III, IV and V elements; in all cases, we will give results for the energy of the first s-orbital bound to the closed-shell core . Next, Section IV presents a simple method for separating valence electron densities from the closed-shell core densities. A rearrangement of the valence electrons' kinetic energy then shows cancellation of the core potential with the kinetic energy density of the valence electrons; a residual valence kinetic energy in the core region remains. Appendix B shows that both the kinetic cancellation and the residual kinetic energy in the core region can be derived from an envelope function approximation for the valence orbitals. Section V demonstrates how to take advantage of these effects in the core region by deriving an effective potential, as well as a total energy equation that depends on either the valence density, in a density functional approach, or, alternatively, the low spatial frequency (LSF) envelopes of the valence orbitals. In Section VI, we present a method for setting the parameters of the effective potential. In Section VII, we bring all the pieces together and present our band structures resulting from the self-consistent Thomas-Fermi valence density and potential on the zinc-blende lattice. We will calculate band structures for most of the III-V and Group IV materials. Section VIII contains our conclusions, along with a brief discussion relating these methods to the rich history of electronic structure calculations.

## II. Revisiting the Thomas-Fermi Model: Lowest orbital corrections

We advocate using a slightly modified Thomas-Fermi density given by

$$\rho(r) = \frac{(2m(F-V(r)))^{3/2}}{3\pi^2 \hbar^3} - \frac{(2m(E_0-V(r)))^{3/2}}{3\pi^2 \hbar^3} + |\Phi_{00}(r)|^2$$
$$\equiv f(F-V(r)) - f(E_0-V(r)) + |\Phi_{00}(r)|^2$$
(2)

in which we define the density function for potential $V(r)$ up to Fermi level $F$ as $f(F-V(r)) \propto (F-V(r))^{3/2}$; note that this density function is zero when $F-V(r) \leq 0$. Also, $\Phi_{00}(r)$ is the lowest energy 1s-orbital at energy $E_0$. This formula is derived in Appendix A.

Equation (2) differs from the standard result in several ways. First, the modified density results from a direct application of the WKB approximation to the electron orbitals, suggesting a higher level of accuracy than is motivated by the standard derivations based on a Fermi gas.[3,4] Second, the improved derivation also leads to an obvious separation of closed-shell core and valence electron densities. Third, for atomic problems, this density remains finite at the nucleus, while the standard result diverges. Fourth, when the Poisson potential and the exchange/correlation components of the potential can both be approximated from the electron density, a self-consistent solution using Eq. (2) gives consistently accurate results for a variety of electronic structure problems. In the following sections, we will demonstrate these features with applications to both ionization potentials and band structure calculations.

**III. Predicting Ionization Potentials for Closed-Shell Ions**

As a first application of the modified Thomas-Fermi, we will calculate the third ionization potential of the Group III elements, the fourth ionization potential of the Group IV elements and finally, the fifth ionization potential of the Group V elements. In all cases, we will calculate the modified density and self-consistent potential for the closed-shell ion, and then solve the radial wave equation for the 'ns'-orbital electron energy and wave function using the self-consistent potential. This approach neglects any influence of the lowest-energy valence electron on the closed-shell ion core and is, therefore, approximate. However, the final results are in reasonable agreement with the measured ionization potentials.

We will detail our calculation for $Si$ with atomic number $Z=14$. We need to find the energy, $I(4)$, required to remove the outer electron from $Si^{+3}$, so that
$Si^{+3} + I(4) \rightarrow Si^{+4} + e$.

First, we calculate the modified Thomas-Fermi density and potential for the ten electrons in $Si^{+4}$. The total potential is given as

$$V = V_P + V_{exc} \quad , \tag{3}$$

in which $V_P$ satisfies the Poisson equation in the radial coordinate as

$$\nabla^2\left(V_P + \frac{Ze^2}{r}\right) = \frac{1}{r}\frac{d^2}{dr^2}\left(r\cdot\left(V_P + \frac{Ze^2}{r}\right)\right) = -4\pi e^2 \rho(r) \quad , \tag{4}$$

and the electron density is the modified density. Here, we pick the lowest-orbital as

$$|\Phi_{00}(r)|^2 = \frac{\alpha^3}{\pi\cdot a_0^3}\exp(-2\alpha r/a_0) \quad , \tag{5}$$

in which $\alpha = Z - 5/16$ and $E_0 = -(Z-5/16)^2$ Rydbergs; this is the standard shielding result for the inner two electrons while neglecting all others.[3] Actually, the results for valence electrons are insensitive to the lowest-orbital shielding estimate, but this guess seems reasonable.

The exchange/correlation parts, $V_{exc}$, are approximated in the local density approximation (LDA) as the derivative of the exchange/correlation energy density

$$V_{exc} \equiv \frac{dU_{exc}}{d\rho} = -(3/\pi)^{1/3}\cdot e^2 \rho(r)^{1/3} \quad , \tag{6}$$

in which we have neglected correlation effects.[1,2] This functional form can be motivated from the "Fermi hole" that each electron forms and carries with it in the presence of parallel spin electrons. The idea of this local approximation for exchange is due to Slater, who derived a formula that was 3/2 times Eq. (6). The correction was obtained by Kohn and Sham.[3]

We solve Eqs. (2) through (6) with an iterative procedure. First, from previous values of $V_{exc}$, we use a predictor-corrector integrator to solve Eq. (4) for the updated Poisson potential on a radial grid. We then adjust the Fermi level and repeat the integration until the volume integral of the density converges to $Z-v=10$ for $Si^{+4}$, in which $v$ is equal to the valence. Next, we use these converged values to find new estimates for $V_{exc}$. We then return to the first step and iterate to overall convergence. The final output of this procedure is the electron density, $\rho$, the potentials, $V_P$ and $V_{exc}$, of the closed-shell ion core, as well as the radius, $R_{ion}$, defined by the radial value at which the self-consistent electron density falls to zero. We construct the ion core potential and continuously connect it to the outer Coulomb potential as

$$V_c(r) = V_P(r) - V_P(R_{ion}) + \kappa \cdot V_{exc}(r) - \frac{v \cdot e^2}{R_{ion}} \qquad r \leq R_{ion}$$
$$V_c(r) = -\frac{v \cdot e^2}{r} \qquad r \geq R_{ion}$$
. (7)

This should be an approximation to the ion core potential seen by the valence electrons. Here, we have included a constant factor, $\kappa$, in order to make adjustments in the strength of valence electron interaction with the ion core LDA exchange potential. $\kappa$ can be simply treated as an adjustable parameter; however, we can make a factor less than unity plausible: In the exchange contribution to the Hartree-Fock eigenvalue for a plane-wave orbital, we encounter a factor $\kappa$ given as

$$\kappa(r) \approx 1 + \frac{1-\eta^2}{2\eta} \log\left|\frac{1+\eta}{1-\eta}\right| \qquad , \qquad (8)$$

in which $\eta(r)$ is the local ratio of the electron momentum to the Fermi momentum for the local electron density.[2,3] In keeping with the Thomas-Fermi approximation, since a valence orbital has higher energy than the core electrons, it becomes clear that $\eta > 1$ and, therefore, $\kappa < 1$ for the core region. We use a constant $\kappa$ in Eq. (7) to roughly mimic this behavior.

Next, we calculate the energy eigenvalue, $\varepsilon$, and orbital eigenfunction, $\psi \equiv u/r$, of the s-orbital bound to the closed-shell ion core potential by using a predictor-corrector integrator on the radial wave equation ($l=0$)

$$\frac{d^2 u}{dr^2} = \left( \frac{l(l+1)}{r^2} + \frac{2m}{\hbar^2} V_c(r) - \frac{2m}{\hbar^2} \varepsilon \right) u \quad . \tag{9}$$

We iterate the radial integrations and adjust the energy eigenvalue until the 'ns'-orbital eigenfunction converges at large radii.

Table I presents our s-orbital eigenvalues for the Group III, IV and V closed-shell ions. Since Eq. (8) suggests that the valence electron might respond to less of the core exchange potential, we show results for two values of $\kappa=.5$ and $\kappa=1$. We also show the experimental results for these ionization potentials. With the exception of the fourth row ions, $Ga, Ge, As$, the values calculated for the range $.5 \leq \kappa \leq 1$ bracket the experimental results. We would obtain better fourth-row results with $\kappa$ slightly less than .5. As expected, the Group IV 'ns'-orbital eigenfunctions for $C^{+3}, Si^{+3}, Ge^{+3}, Sn^{+3}$ and $Pb^{+3}$ show 2s, 3s, 4s, 5s and 6s character respectively. Perhaps, if we used the more exact non-local form of exchange potential in the radial wave equation, or allowed the core to be perturbed by the valence electron, we could improve these LDA results. Despite these approximations, it appears that the modified Thomas-Fermi density provides reasonably accurate descriptions of the closed-shell ion cores for Groups III, IV and V of the periodic table. These results will be useful later in Section VI.

**IV. Separating Core from Valence Electrons: Kinetic and Potential Cancellation**

One extremely convenient feature of the Thomas-Fermi density is the facile separation of valence electrons from core electrons. Consider rewriting Eq. (2) as

$$\rho(r) = f(F - V(r)) - f(F_c - V(r))$$
$$+ f(F_c - V(r)) - f(E_0 - V(r)) + |\Phi_{00}(r)|^2 \quad , \tag{10}$$

in which we can identify the first two terms as the valence density, $\rho_v$, and the last three terms as the closed-shell core density, $\rho_c$. Here, the new parameter is the Fermi level setting for the core, $F_c$; as always, it is adjusted to fix the number of core electrons.

We can now calculate the kinetic energy density of these valence electrons by following a procedure similar to that mapped out in Appendix A. The familiar final result is

$$t_v = \frac{3}{5}(F_c + \delta F - V(r))f(F_c + \delta F - V(r)) - \frac{3}{5}(F_c - V(r))f(F_c - V(r)) \quad r \leq r_c$$
$$= \frac{3}{5}(F_c + \delta F - V(r))f(F_c + \delta F - V(r)) \quad r > r_c$$
$$\tag{11}$$

in which we redefine the upper Fermi level as $F \equiv F_c + \delta F$, so that the valence electrons occupy an energy range, $\delta F$. Also, we define the core region radius at $F_c - V(r_c) = 0$; we will have much more to say about $r_c$ in the next sections. This valence kinetic energy density can be rearranged in a useful way inside the atomic core region, $r \leq r_c$. Consider the exact rearrangement

$$t_v(r) = (F_c - V_c(r))\rho_v + \int_0^{\delta F} ds(s - V_v)\frac{df(s - V_v + F_c - V_c)}{ds} \quad r \leq r_c \quad , \tag{12}$$

in which $V_c$ is the potential (direct, exchange and correlation) of the ion core, while $V_v$ is the potential due to the valence electrons. The first term in this kinetic energy density, when added to the valence potential energy density, will give a perfect cancellation of the ion core potential. The second term, containing the valence electron density of states, $df/ds \equiv d\rho/ds$, can be interpreted as a residual kinetic energy density for the valence electrons responding only to the valence potential; recall that $s$ is the increment in energy above the Fermi level of the core electrons.

The kinetic energy density cancellation in the ion core region can also be developed from the valence orbitals. Consider Hartree-Fock valence orbitals satisfying the equations

$$H\psi_n = \varepsilon_n \psi_n \equiv (T + V_c + V_v)\psi_n \qquad n_c < n \leq M \qquad . \qquad (13)$$

Here, we can identify the Fermi level for the ion core, $F_c \approx \varepsilon_{n_c}$, as well as the highest valence level, $F_c + \delta F \approx \varepsilon_M$. Using all the valence orbital equations, we can easily construct the kinetic energy density of the valence electrons and rearrange terms as

$$t_v(r) = (F_c - V_c(r))\sum_{n_c+1}^{M} \psi_n \cdot \psi_n^* + \sum_{n_c+1}^{M}(\varepsilon_n - F_c - V_v(r))\psi_n \cdot \psi_n^* \qquad r \leq r_c \qquad . \qquad (14)$$

The first term suggests cancellation of valence kinetic and ion core potential energy densities and is obviously equivalent to the first term in Eq. (12). The second term corresponds to the residual kinetic energy density on the right-hand-side of Eq. (12). In the core region, this requires

$$\sum_{n_c+1}^{M}(\varepsilon_n - F_c - V_v)\psi_n \cdot \psi_n^* \approx \int_0^{\delta F} ds(s - V_v)\frac{d\rho}{ds} \qquad r \leq r_c \qquad . \qquad (15)$$

We can establish Eq. (15) by using a WKB approximation for the orbitals, $\psi_n$, while converting the sum on the left-hand-side to an integral over the valence energy range, $\delta F \approx \varepsilon_M - F_c$; the relevant manipulations are demonstrated in Appendix A.

## V. Unperturbed Cores and Valence-Only Equations

The relation shown in Eq. (12) can greatly simplify all valence electron calculations. In particular, if we fix the potential of the closed-shell ion cores, $V_c$, assuming that the core electrons are not greatly influenced by the rearrangements of the valence electrons, then for an atomic problem, the total energy of the valence electrons is given by the sum of the kinetic and potential energies as

$$
\begin{aligned}
U_v &= \int d^3 r \, t_v(r) + \int d^3 r (V_c(r)\rho_v + U_{exc}(\rho_v)) + \frac{1}{2}\iint d^3 r d^3 r' \frac{\rho_v(r)\rho_v(r')}{|\vec{r}-\vec{r}'|} \\
&\equiv T_{eff} + \int d^3 r (V_{eff}(r)\rho_v + U_{exc}(\rho_v) + F_c \rho_v) + \frac{1}{2}\iint d^3 r d^3 r' \frac{\rho_v(r)\rho_v(r')}{|\vec{r}-\vec{r}'|}
\end{aligned} \quad (16a)
$$

Here, we have used the LDA form of exchange/correlation. The residual valence electron kinetic energy in the core region, along with the kinetic energy outside the core region, have been combined as

$$
T_{eff} \equiv \int_{r \leq r_c} d^3 r \int_0^{\delta F} ds (s - V_v) \frac{d\rho}{ds} + \int_{r > r_c} d^3 r \frac{3}{5}\frac{\hbar^2}{2m}(3\pi^2)^{2/3}\rho_v^{5/3} \quad . \quad (16b)
$$

Furthermore, the complete kinetic cancellation of the core potential, due to the first term in Eq. (12), leads to an effective potential, $V_{eff}$, for the closed-shell ion. For an ion with valence, $v$, we find

$$
\begin{aligned}
V_{eff}(r) &\equiv 0 & r \leq r_c \\
&\equiv -\frac{v \cdot e^2}{r} & r > r_c
\end{aligned} \quad . \quad (16c)
$$

The second expression for $U_v$ suggests that the valence electrons are almost free particles in the inner core region, thanks to kinetic cancellation. Results of this sort were first presented by Ashcroft in 1966.[6,7,8] The details of the ion core potential that were obviously critical for the ionization potential calculations in Section II seem to have totally disappeared. Actually, this is somewhat of an illusion, since the detailed core potential influences $V_{eff}$, $T_{eff}$ and $U_v$ in several ways. First, there is the valence density of states, $d\rho/ds$. Second, there is the core radius, $r_c$. This is defined by the equation, $F_c - V(r_c) = 0$, in which the total self-consistent potential generated by both the ion core and valence electrons appears. Third, the core Fermi level, which is fixed by the number of core electrons, also depends on $V(r)$ through the core density expression. Finally, since an unperturbed core is just an approximation, is it useful to fix these core parameters while calculating a self-consistent valence density that minimizes the valence energy $U_v$? We will assume that this is a decent approximation, and that once we arrive at a minimum energy valence density, we can, if necessary, improve and iterate on the ion cores as perturbed by the rearrangements of the valence electrons.

We proceed toward minimizing the valence energy by, first, approximating the expression for the residual kinetic energy in the region $r \leq r_c$. Since $V_{eff} = 0$ in the core region, the residual valence kinetic energy is $(s - V_v)$. In keeping with the Thomas-Fermi approximation, we define a LSF valence density in the core region as

$$(s - V_v) \equiv \frac{\hbar^2}{2m}(3\pi^2)^{2/3}\rho^{2/3} \qquad r \leq r_c \quad . \tag{17}$$

This LSF density is related to the slowly varying envelope of the exact valence density. The next approximation is critical: In the core region, we replace the exact valence density with the LSF valence density. Appendix B develops this approximation, showing that the LSF density can be derived from the slowly varying envelope of the valence orbitals. Finally, we calculate the residual core region kinetic energy term as

$$\int\limits_{r \leq r_C} d^3r \int\limits_0^{\delta F} ds(s-V_v)\frac{d\rho}{ds} \approx \int\limits_{r \leq r_C} d^3r \int\limits_0^{\rho_v} d\rho \frac{\hbar^2}{2m}(3\pi^2)^{2/3}\rho^{2/3} = \int\limits_{r \leq r_C} d^3r \frac{3}{5}\frac{\hbar^2}{2m}(3\pi^2)^{2/3}\rho_v^{5/3} \ .$$

(18)

With the Eq. (18) result, the valence kinetic energy expression from outside the core is carried into the core region as well. Now, we obtain our final expression for the total valence energy as

$$U_v - F \cdot N_v = \int d^3r \frac{3}{5}\frac{\hbar^2}{2m}(3\pi^2)^{2/3}\rho_v^{5/3} + \int d^3r (V_{eff}(\vec{r})\rho_v + U_{exc}(\rho_v))$$
$$+ \frac{e^2}{2} \iint d^3r d^3r' \frac{\rho_v(\vec{r})\rho_v(\vec{r}')}{|\vec{r}-\vec{r}'|} - \delta F \int d^3r \rho_v$$

(19)

Here, in anticipation of minimizing the valence energy while maintaining a fixed number of valence electrons, $N_v$, we have introduced a Lagrange Multiplier, $F$; notice that the final combination on the right-hand-side is just what one would expect for a valence-only Fermi level measured relative to the core Fermi level, $\delta F \equiv F - F_c$. Ultimately, as we discuss later, we adjust the increment in Fermi level, $\delta F$, to fix the number of valence electrons. Equation (19) is the final embodiment of kinetic cancellation and the unperturbed core assumption, while allowing for valence envelope degrees of freedom. While the core electron density and potential do not appear, their effects are present in the effective potential, $V_{eff}$. Also, as shown in Appendix B, should we want to improve on the envelope function approximation to the core valence structure, we can incorporate the core potential-dependent oscillations shown in Eq. (1B). Although we wrote Eq. (19) for an atomic problem, it readily generalizes to molecular and solid-state calculations, where we must include multiple ion cores, or an appropriate lattice of ion cores, as well as their mutual interactions. Note that these ion core interactions do depend on the detailed distributions of the core electrons, while for higher accuracy, the high spatial frequency part of the valence densities could be treated as a small perturbation. Finally, we have assumed that when we apply the formalism to molecular and solid-state problems, the valence density will generalize to a function of $\vec{r} = (x, y, z)$.

At this point, we can generate an LSF orbital version of Eq. (19). If we identify the valence density as a sum over LSF valence orbital densities, $\{\bar{\psi}_n^* \bar{\psi}_n\}$, as developed in Eq. (5B) of Appendix B, we can rewrite the valence energy as

$$U_v = \sum_n \int d^3r\, \bar{\psi}_n^* \left( -\frac{\hbar^2}{2m} \nabla^2 - V_{eff}(\vec{r}) \right) \bar{\psi}_n \\ + \frac{e^2}{2} \iint d^3r\, d^3r' \frac{\sum_{n<l} \bar{\psi}_n^* \bar{\psi}_n \sum_l \bar{\psi}_l^* \bar{\psi}_l}{|\vec{r} - \vec{r}'|} + \int d^3r\, U_{exc}\left( \sum_n \bar{\psi}_n^* \bar{\psi}_n \right)$$

. (20)

If we take the stationary variation of Eq. (20) and impose orthonormality of the orbitals as an auxiliary condition, we can then obtain equations very much like Hartree-Fock equations for the valence orbital functions, $\{\bar{\psi}_n\}$. Note that these valence orbitals match the exact valence orbitals outside the core region, while inside the core region they represent the slowly varying envelopes of the true valence orbitals. Equation (20), supplemented by an effective potential due to multiple ion cores, could provide an excellent starting point for numerical chemistry calculations in the LDA.

## VI. Estimating the Core Radius

To start a calculation, whether atomic, molecular or solid-state, we need to fix the valence, $v$, as well as estimate the core radius, $r_c$, for each elemental closed-shell ion in the problem. These two parameters then allow us to specify the effective potential, $V_{eff}$, for each ion. Here, we explain one method to obtain an initial estimate of the effective potential for each of the ionized Group III, IV and V elements.

As originally advocated by Ashcroft in 1966, we use the ionization potentials calculated in Section III.[8] We proceed by calculating the energy eigenvalue, $\bar{\varepsilon}$, and LSF orbital eigenfunction, $\bar{\psi} \equiv \bar{u}/r$, of the s-orbital bound to the effective potential in the radial wave equation,

$$\frac{d^2 \bar{u}}{dr^2} = \left( \frac{2m}{\hbar^2} V_{\mathit{eff}}(r) - \frac{2m}{\hbar^2} \bar{\varepsilon} \right) \bar{u} \quad . \tag{21}$$

This equation results from a stationary variation of the first line in Eq. (20); obviously, valence self-interaction and exchange cancel when there is but one valence electron. We choose $v = 3, 4, 5$ for Group III, IV and V ions respectively, and then make an initial guess for the core radius. Next, we use a predictor-corrector algorithm to solve Eq. (21) and adjust the energy eigenvalue until the eigenfunction approaches zero at large radii. We then make adjustments in the core radius, $r_c$, until the eigenvalue, $\bar{\varepsilon}$, is equal to the ionization potential, $\varepsilon$, in Table I. Recall that the calculated $\varepsilon$ was obtained by integrating Eq. (9) for the full ion core potential. Figure 1 shows the calculated 5s wave function for the electron bound to the closed shell antimony ion, $Sb^{+5}$, as well as the degenerate solution to Eq. (21); note the exact match for the region $r > R_{ion}$. Continuing with the method based on measured ionization potentials, we calculate the $\tilde{r}_c$ values in Table II. For all Group III, IV and V elements, the true value is close to but normally smaller than $\tilde{r}_c$.[8] The core radii in the last column of Table II will be used for the band diagram calculations in Section VII.

## VII. Band Structure Calculations

Every electronic structure calculation proceeds by minimizing the total valence energy with respect to the envelope of the valence density and then solving the resulting equations. When we set the valence density variation of Eq. (19) equal to zero, we find

$$\frac{\hbar^2}{2m}(3\pi^2)^{2/3}\rho_v(\vec{r})^{2/3} + V_{eff}(\vec{r}) + V_{exc}(\vec{r}) + e^2\int d^3r' \frac{\rho_v(\vec{r}')}{|\vec{r}-\vec{r}'|} = \delta F \quad , \tag{22}$$

in which the valence Fermi level increment, $\delta F$, is adjusted to fix the number of valence electrons. The formal solution for the valence density is

$$\rho_v(\vec{r}) = \frac{(2m(\delta F - V_{eff} - V_P - V_{exc}))^{3/2}}{3\pi^2 \hbar^3} \quad , \tag{23a}$$

in which the Poisson potential satisfies

$$\nabla^2 V_P = -4\pi e^2 \rho_v(\vec{r}) \tag{23b}$$

while the LDA approximation for exchange and correlation gives

$$V_{exc} = -(3/\pi)^{1/3} \cdot e^2 \rho_v(\vec{r})^{1/3} - \frac{e^2}{a_0}(.0311 \cdot \log(a_0 \rho_v(\vec{r})^{1/3}) + .07322) \quad . \tag{23c}$$

Here, we have included a correlation correction. For band calculations, the correlation correction can change the bandgaps and other features by about ten percent. The correlation energy whose variation leads to this functional form for potential is discussed in detail in Reference 9. $V_{eff}$ is given in Eq. (16c) calculated with appropriate values for valence and the core radius. Once Eqs. (23a), (23b) and (23c) are solved to convergence, we have self-consistent values for both the valence density, $\rho_v(\vec{r})$ and the total potential sensed by the valence electrons, $V_T = V_{eff} + V_P + V_{exc}$.

Equations (23a) to (23c), with kinetic cancellation manifested in the effective potential of the ion cores, are the starting points for electronic structure calculations using the Thomas-Fermi method. All of the derivations and manipulations have led us to this point, and we are now ready to attack some important problems, such as the band structure of crystalline solids. The key feature of this problem set is the periodic arrangement of the closed-shell ions on a lattice. We will assume that the lattice structure is known, and we will solve for the periodic valence density as well as the total periodic potential felt by the valence electrons. The band structure follows directly from this periodic potential. We could, if desired, include the mutual interactions of the closed-shell ion cores in Eq. (19), and then find the lattice configuration that minimizes total energy. Recall that the detailed charge densities of these cores, needed to calculate the interactions, were well approximated in Section III.

When lattice periodicity is present in a problem, taking some of the calculation into reciprocal space has advantages. For what follows, we will specialize to the zinc-blende lattice with lattice constant, $a$. When we transform the total lattice valence potential into the reciprocal lattice space, $\{\vec{g}\}$, we find the relation

$$V_T(\vec{r}) = V_{eff} + V_P + V_{exc} = \sum_g (\tilde{V}_{eff} + \tilde{V}_P + \tilde{V}_{exc}) e^{i\vec{g}\cdot\vec{r}} = \sum_{\vec{g}} \tilde{V}_T(\vec{g}) e^{i\vec{g}\cdot\vec{r}} \quad , \tag{24}$$

in which the Fourier transform of the effective potential of the ion core lattice is given as

$$\tilde{V}_{eff}(\vec{g}) = \frac{-4\pi e^2}{\Omega(\vec{g}\cdot\vec{g})} (v_c \cos(g\, r_c^c) e^{-i\vec{g}\cdot\vec{s}} + v_a \cos(g\, r_c^a) e^{i\vec{g}\cdot\vec{s}}) \quad . \tag{25}$$

The primitive cell volume is $\Omega$, while $2\cdot\vec{s} = (a/4, a/4, a/4)$ is the basis vector between the cation and anion with valences $(v_c, v_a)$ and core radii $(r_c^c, r_c^a)$ respectively. Similarly, the Poisson potential on the lattice is given by the transform of the valence density as

$$\tilde{V}_P(\vec{g}) = \frac{4\pi e^2}{\Omega(\vec{g}\cdot\vec{g})} \tilde{\rho}_v(\vec{g}) \quad . \tag{26}$$

We solve Eqs. (23a) and (23b) iteratively as follows: 1) Using Eq. (23a) with the total potential, $^{old}V_T(\vec{r})$, evaluated in a unit cell of the lattice, we make adjustments in the Fermi level, $\delta F$, until the valence density integrates to $(v_a + v_c)$ valence electrons per unit cell; 2) We calculate the lattice space exchange/correlation potential and then transform it to reciprocal space, $\tilde{V}_{exc}(\vec{g})$; 3) We then transform the valence density to reciprocal space and solve Eqs. (25) and (26) for the sum $\tilde{V}_{eff}(\vec{g}) + \tilde{V}_P(\vec{g})$, while, for charge neutrality, we fix the $\vec{g} = 0$ term to zero; 4) We form the total potential in reciprocal space, $\tilde{V}_T(\vec{g})$, and use linear mixing to estimate the new total potential for the next iteration as $^{new}\tilde{V}_T(\vec{g}) \equiv \tilde{V}_T(\vec{g}) \cdot \beta + {}^{old}\tilde{V}_T(\vec{g}) \cdot (1 - \beta)$.[7] We force the new potential to zero beyond a cut-off in reciprocal space, and then update the old potential to the new potential; 5) We transform the potential back to lattice space. This completes one iteration. We typically set $\beta = .4$, and we stop the interations when the lattice potential and valence density have converged.

Once the total lattice potential is converged, we solve the reciprocal space wave equation for the bands. Bloch's theorem for the slowly varying envelopes of the valence states is satisfied by the form

$$\overline{\psi}_n(\vec{r}) = e^{i\vec{k}\cdot\vec{r}} \sum_{\vec{g}} b_n(\vec{g},\vec{k}) e^{i\vec{g}\cdot\vec{r}} \quad , \tag{27}$$

in which $\vec{k}$ is the Bloch momentum. When we neglect the spin-orbit interaction (see Appendix B), the wave equation in reciprocal space is then

$$\left(\frac{\hbar^2}{2m}\left|\vec{k} + \vec{g}\right|^2 - E_n(\vec{k})\right) b_n(\vec{g},\vec{k}) + \sum_{\vec{g}'} \tilde{V}_T(\vec{g} - \vec{g}') b_n(\vec{g}',\vec{k}) = 0 \tag{28}$$

in which $b_n(\vec{g},\vec{k})$ are the reciprocal space coefficients for the Bloch function and $E_n(\vec{k})$ give the band structure. Our band structure results for nine III-V materials are given in Figs. 2a to 2i. In all cases, we set the cation valence to three and anion valence to five, using the core radius values given in the last column of Table II. Similar results for the Group IV diamond structures are in Figs. 3a to 3c. For all of the materials, we used a cut-off for the potential in reciprocal space at $\vec{g}\cdot\vec{g} \leq 16(2\pi/a)^2$.

The band structures for the binary semiconductors are quite good, given that we have not included the spin-orbit interaction. The direct bandgap materials, GaAs, InAs, InP, GaSb and InSb, are found to be direct, while the predicted band gaps and other features are fairly accurate. Similarly, the indirect bandgap materials, AlAs, AlSb, GaP and AlP, are calculated to be indirect with reasonable accuracy on other band features. In all cases, we expect that slight adjustments in the ion core radii and, of course, the spin-orbit interaction will affect the details. A fairly accurate method for including the spin-orbit interaction in reciprocal lattice space is discussed in Reference 11. When we look to the Group IV band diagrams, we need to point out another sensitivity in the band features. If we reduce the LDA exchange potential to .85 times the Kohn and Sham version, then all of the band features become more accurate. The results for Si, Ge and Sn were all calculated using this slightly reduced value for the exchange potential. Furthermore, these band results illustrate an extremely important feature: the parameter values that define the effective potential of the ion cores are transferable. For example, the parameters that define the gallium ion core are used in GaP, GaAs, as well as GaSb.

## VIII. Conclusions

The modified Thomas-Fermi approximation as given in Eqs. (2), (3) and (4) retains a high level of accuracy for the ion core potential and should work as well for a large variety of atomic physics problems. The modified form of the density avoids many of the Thomas-Fermi shortcomings, such as infinite density near the nucleus and infinite radius for the neutral atom.[3] Our Table I results for the third, fourth and fifth ionization potentials show good agreement with the experimental data, particularly when we include a factor, $\kappa \leq 1$, to adjust the strength of the ion core LDA exchange potential. We can easily motivate this factor by considering the exact exchange contribution to the Hartree-Fock eigenvalue for a plane-wave orbital.[3] Finally, we must remember that these calculations are necessarily approximate, as they neglect any influence of the single valence electron on the self-consistent closed-shell ion cores.

Next, we separated Thomas-Fermi core electron densities from valence electron densities by introducing the core Fermi level, $F_c$. Then, when we calculated the valence kinetic energy density, we showed how it separated exactly into two terms, the first of which canceled the potential of the ion cores in the core region, while the second represented the residual kinetic energy driven by the LSF valence density. Furthermore, Appendix B detailed how these terms resulted from a slowly varying envelope approximation to the exact valence orbitals. This kinetic cancellation allowed us to write a functional for the total valence energy, in which the effective potential, $V_{eff}$, replaced the ion core potential. These processes of kinetic cancellation of the strong ion core potential, as well as replacing the valence density in the core with a LSF valence density, are the most critical and beneficial steps in the entire procedure. That kinetic cancellation might occur in a Thomas-Fermi based theory was first suggested in the early days of the development of pseudopotentials.[6] These early presentations expanded the traditional Thomas-Fermi kinetic energy density as

$$\frac{3}{5}\frac{\hbar^2}{2m}(3\pi^2)^{2/3}(\rho_c+\rho_v)^{5/3} \approx \frac{3}{5}\frac{\hbar^2}{2m}(3\pi^2)^{2/3}\rho_c^{5/3} + \frac{\hbar^2}{2m}(3\pi^2)^{2/3}\rho_c^{2/3}\rho_v + \ldots \quad ; \qquad (29)$$

the second term in this expansion leads to kinetic cancellation. This procedure was limited, as it gave no indication of how to include the higher-order terms and was only useful when $(\rho_v/\rho_c)$ was a small quantity. The derivations that we gave in Sections IV, V and Appendix B removed this restriction, yielding both the exact core potential cancellation and the residual kinetic energy of the valence electrons in the core region. Also, whenever higher accuracy is needed, the high-spatial-frequency part of the valence densities in the core regions could be incorporated as a small perturbation. This kinetic cancellation, based on a slowly varying envelope approximation for the valence orbitals, differs from the standard approach that emphasizes orthogonality of the valence and core electron orbitals.[2]

In order to implement the equations for the LSF valence density, we needed to develop a procedure to set the valence, $v$, and the core radius, $r_c$, for each elemental closed-shell ion in the problem. In Section VI, we discussed an estimate based on measured or calculated ionization potentials; in all cases, this estimate for the core radius provided an upper bound as shown in Table II. We would certainly like to improve on this estimate to get us nearer to the values in the final column of Table II. Perhaps, in a future method, we might return to the equation, $F_c - V_c(r_c) - V_v(r_c) = 0$, in which the total self-consistent potential, as well as the core Fermi level, appear.

In Section VII, we minimized the total valence energy resulting in the valence density solution in Eqs. (22) and (23). These valence-only equations provide a basis for atomic, molecular and solid-state electronic structure calculations. Here, we used them to calculate the band structures resulting from the self-consistent valence density and potential on the zinc-blende and diamond lattices. Our band structure results for most of the Group III-V and Group IV semiconductor materials indicate that there is a remarkable level of reality present in these equations.[12] Also, it appears that the ion cores, as defined by the effective potential of Eq. (16c), are completely transferable among material systems. The parameters that define the gallium ion core are used in GaP, GaAs, as well as GaSb, and so on for all the ion cores. This feature is very important for all implementations of the method.

Why is the Thomas-Fermi approximation working well in these applications? The critical element is the elimination of high potential gradients and the resulting dependence of the energy solely on low spatial frequency effective valence potentials and valence densities. In this regard, the results of Appendix B on the valence envelope approximation are absolutely crucial. It does not particularly matter whether we base our valence electron structure calculations on density with Eq. (19) or on orbitals with Eq. (20); they both can provide excellent approximations for the LSF valence structure. However, whenever gradients of potential are relatively low, the Thomas-Fermi approximation has always offered decent accuracy, and that is the case here.

## Appendix A: A Modified Thomas-Fermi Density

We will work a one-dimensional example first. Consider a density for N electrons given in terms of orthogonal orbitals as

$$\rho(x) = 2 \cdot \sum_{n=0}^{M} |\Phi_n(x)|^2 \quad , \tag{1A}$$

in which we have placed a spin-up and spin-down electron into each orbital from the lowest energy, n=0, to the highest occupied level, n=M. When we approximate the modulus squared of each orbital using the Wentzel-Kramers-Brillouin method, WKB, we find

$$\rho(x) = \sum_n N_n \cdot \frac{2}{\sqrt{2m(E_n - V(x))}} \equiv \sum_n N_n \cdot \frac{2}{p(x, E_n)} \quad , \tag{2A}$$

in which the orbital energies are given by $E_n$ and $N_n$ is a normalization factor. Also, we have spatially averaged the $\left\langle \sin^2 \int^x dx' p(x', E_n)/\hbar \right\rangle \approx 1/2$ factor in each term. If we differentiate the quantization condition,

$$2 \cdot \int dx \sqrt{2m(E_n - V(x))} = (n + 1/2) \cdot h \quad , \tag{3A}$$

with respect to n, we find that the normalization is given by $2m/h \cdot \frac{dE_n}{dn}$ and

$$\rho(x) = \sum_n 2m/h \cdot \frac{dE_n}{dn} \frac{2}{p(x, E_n)} \approx \int_{E_0}^{F} dE \frac{4m/h}{p(x, E_n)} + |\Phi_0(x)|^2 \quad . \tag{4A}$$

Here we have replaced the sum with an integral supplemented by the modulus squared of the lowest orbital; these are low-order terms in the Euler-Maclaurin formula.[5] Also, we set the lower limit of the integral to $E_0$ and the upper limit to the Fermi energy, F; this should be approximately the energy for the highest occupied level. In applications, F is always adjusted to give the correct number of electrons. Finally, we have

$$\rho(x) = 4/h \cdot (\sqrt{2m(F-V(x))} - \sqrt{2m(E_0-V(x))}) + |\Phi_0(x)|^2 \quad . \tag{5A}$$

Only the first term, with no lowest-orbital corrections, would be present in the standard one-dimensional Thomas-Fermi result.

The density derivation for a spherically symmetric potential follows a similar development from a slightly more complicated starting point. Consider a density for N electrons given in terms of orthogonal orbitals in spherical coordinates as

$$\rho(r,\theta,\phi) = 2 \cdot \sum_{n=0}^{M} \sum_{l} \sum_{m=-l}^{l} |\Phi_{nl}(r)/r|^2 \, Y_{lm}(\theta,\phi) \cdot Y_{lm}^*(\theta,\phi) \quad , \tag{6A}$$

in which $\Phi_{nl}(r)$ is a solution of the radial wave equation. With the addition theorem for the spherical harmonics, we simplify this to a radial density

$$\rho(r) = 2 \cdot \sum_{n=0}^{M} \sum_{l=0}^{} |\Phi_{nl}(r)/r|^2 \, \frac{2l+1}{4\pi} \quad . \tag{7A}$$

We can now follow the development of the one-dimensional example, using the WKB approximation for the solutions to the radial wave equation, as well as the normalization obtained from the radial quantization condition. This gives

$$\rho(r) = \frac{2}{4\pi} \cdot \sum_{n=0}^{M} \sum_{l} \frac{dE_{nl}}{dn} \frac{2m}{\hbar^2} \frac{d\beta}{dl} \frac{2m}{h \cdot \sqrt{2m(E_{nl} - V(r) - \beta)}} \quad , \tag{8A}$$

in which we introduced a quantity $\beta \equiv \frac{\hbar^2 l(l+1)}{2mr^2}$ and $\frac{d\beta}{dl} = \frac{\hbar^2 (2l+1)}{2mr^2}$; note that this last expression is unchanged if $l(l+1)$ is replaced with Langer's correction, $(l+1/2)^2$.[10] Now, as in the one-dimensional case, we replace the discrete sums with integrals. Here, we have two sets of integral limits to set. For the $\beta$-integration, we use limits at $\beta = 0$ and E-V. Next, for the sum over the principle quantum number, we use the leading terms in the Euler-Maclaurin formula. This leads to

$$\rho(r) \approx \frac{2}{4\pi} \cdot \int_{E0}^{F} dE \int_{0}^{E-V} d\beta \frac{2m}{\hbar^2} \frac{2m}{h \cdot \sqrt{2m(E - V(r) - \beta)}} + |\Phi_{00}(r)|^2 \quad . \tag{9A}$$

These final integrations give the modified Thomas-Fermi density

$$\rho(r) = \frac{(2m(F - V(r)))^{3/2}}{3\pi^2 \hbar^3} - \frac{(2m(E_0 - V(r)))^{3/2}}{3\pi^2 \hbar^3} + |\Phi_{00}(r)|^2 \quad . \tag{10A}$$

We discuss the benefits of this formula in the main body of the paper. Primarily, this modified form of the density avoids infinite density near the nucleus and, when exchange is included, infinite radius for the neutral atom.[3]

# Appendix B: The Valence Envelope Function and Density

Consider writing a valence orbital function as

$$\psi_{nlm}(r,\theta,\phi) \equiv \sqrt{2}\sin\left(\int_0^r dr' P_c(r')/\hbar\right)\phi_{nl}(r)Y_{lm}(\theta,\phi)/r \equiv \sqrt{2}\sin(u(r))\phi_{nl}(r)Y_{lm}(\theta,\phi)/r \quad (1B)$$

in which $P_c(r) = \sqrt{2m(F_c - V_c(r))}$ and $du/dr = P_c(r)/\hbar$. We now calculate the kinetic energy density of this valence orbital as

$$\psi_{nlm}^*\left(\frac{-\hbar^2}{2m}\nabla^2\right)\psi_{nlm} = |Y_{lm}(\theta,\phi)/r|^2 \phi_{nl}^* 2\sin^2(u)\left[\frac{-\hbar^2}{2m}\frac{d^2}{dr^2} + (F_c - V_c) + \frac{\hbar^2 l(l+1)}{2mr^2}\right]\phi_{nl}$$
$$+ |Y_{lm}(\theta,\phi)/r|^2 \phi_{nl}^* \sin(2u)\left(\frac{-\hbar^2}{2m}\right)\left[\frac{d^2u}{dr^2} + 2\frac{du}{dr}\frac{d}{dr}\right]\phi_{nl} \quad . \quad (2B)$$

Equation (2B) is exact. However, if we average over radial increments, $\Delta r$, such that $u(r+\Delta r) - u(r) = \pi$, then the first line containing the $2\sin^2(u)$ factor dominates. The second line contributes little on the radial interval, since the average of $\sin(2u)$ is zero, while, if we assume the WKB value, $\phi_{nl} \approx \sqrt{\hbar/P_c}$, then $\left[\frac{d^2u}{dr^2} + 2\frac{du}{dr}\frac{d}{dr}\right]\phi_{nl} \approx 0$ as well.

This averaging, or smoothing, results in a kinetic energy density per orbital in the core region as

$$\psi_{nlm}^*\left(\frac{-\hbar^2}{2m}\nabla^2\right)\psi_{nlm} \cong |Y_{lm}(\theta,\phi)/r|^2 \phi_{nl}^*\left[\frac{-\hbar^2}{2m}\frac{d^2}{dr^2} + (F_c - V_c) + \frac{\hbar^2 l(l+1)}{2mr^2}\right]\phi_{nl} \quad . \quad (3B)$$

At this same level of approximation, the radial interval average of the kinetic and potential energies for the valence orbital gives

$$|Y_{lm}(\theta,\phi)/r|^2 \phi_{nl}^* \left[ \frac{-\hbar^2}{2m}\frac{d^2}{dr^2} + (F_c - V_c) + \frac{\hbar^2 l(l+1)}{2mr^2} + V \right] \phi_{nl} \quad . \tag{4B}$$

In this form, the kinetic cancellation of the core potential and the first appearance of a LSF effective potential is obvious. The envelope function approximation to the orbital is given as

$$\overline{\psi}_{nlm} \approx \phi_{nl}(r) Y_{lm}(\theta,\phi)/r \quad , \tag{5B}$$

while the more exact orbital is given by Eq. (1B). At this point, we could use the LSF orbitals of Eq. (5B) to derive a LSF valence density. Whenever higher accuracy is needed, the high spatial frequency part of the valence densities in the core regions could be treated as a small perturbation.

When the spin-orbit interaction is included, Eq. (4B) is modified in an obvious way by making the valence orbitals Pauli spinors and including the expectation value of the potential term

$$V_{SO} = \frac{1}{2mc^2} \frac{1}{r} \frac{dV}{dr} \vec{S} \cdot \vec{L} \quad . \tag{6B}$$

Here, $\vec{S}$ and $\vec{L}$ are the spin and orbital angular momentum operators. Although kinetic cancellation removes the core potential from the slowly varying envelope equations, its gradient remains in the spin-orbit potential. A method for including the spin-orbit interaction in reciprocal lattice space is discussed in Reference 11.

**Acknowledgments**
This research was funded by the Air Force Research Laboratory's Directed Energy Directorate under contract to Boeing LTS, (Contract Number FA9541-04-D-0402).

by: A. Ongstad, R. Kaspi, G. Dente and M. Tilton in IX International Conference on Mid-Infrared Materials and Devices, 7-11 September 2008 (Freiberg, Germany).

| Element | | I (eV) | K=0.5 | K=1.0 |
|---|---|---|---|---|
| Boron | (B) | 37.93 | 37.15 | 38.30 |
| Aluminum | (Al) | 28.45 | 28.28 | 31.13 |
| Gallium | (Ga) | 30.71 | 32.72 | 38.94 |
| Indium | (In) | 28.02 | 25.66 | 30.46 |
| Thallium | (Tl) | 29.83 | 26.63 | 32.51 |
| Carbon | (C) | 64.49 | 63.45 | 65.02 |
| Silicon | (Si) | 45.14 | 44.63 | 48.01 |
| Germanium | (Ge) | 45.72 | 47.56 | 54.35 |
| Tin | (Sn) | 40.73 | 37.36 | 42.56 |
| Lead | (Pb) | 42.32 | 37.68 | 43.91 |
| Nitrogen | (N) | 97.89 | 96.59 | 98.56 |
| Phosphorus | (P) | 65.02 | 64.14 | 68.03 |
| Arsenide | (As) | 62.63 | 64.35 | 71.70 |
| Antimony | (Sb) | 55.97 | 50.42 | 55.99 |
| Bismuth | (Bi) | 55.97 | 49.48 | 56.40 |

**Table I. Measured and calculated ionization potentials.**

| Element | | $\tilde{r}_c (\text{A})$ | $r_c (\text{A})$ |
|---|---|---|---|
| **Valence = 3** | | | |
| Aluminum | (Al) | .62 | .61 |
| Gallium | (Ga) | .59 | .56 |
| Indium | (In) | .60 | .60 |
| **Valence = 4** | | | |
| Silicon | (Si) | .57 | .53 |
| Germanium | (Ge) | .56 | .51 |
| Tin | (Sn) | .63 | .57 |
| **Valence = 5** | | | |
| Phosphorus | (P) | .51 | .475 |
| Arsenide | (As) | .53 | .47 |
| Antimony | (Sb) | .61 | .53 |

**Table II. Estimated core radii, $\tilde{r}_c$, and core radii, $r_c$, used in band calculations for all Group III, IV and V elements.**

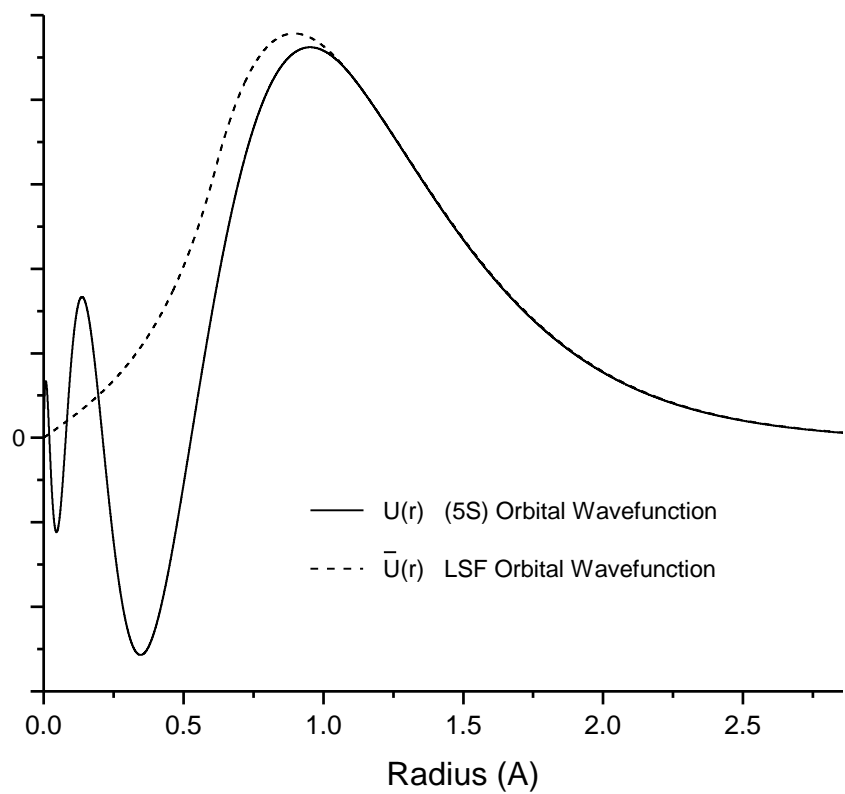

**Fig. 1. Calculated orbital (5S) and degenerate low spatial frequency orbital for a valence electron bound to $Sb^{+5}$.**

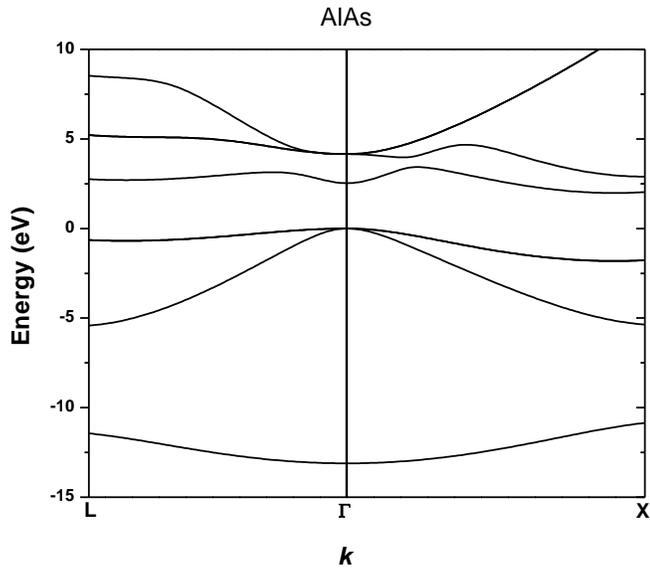

**Fig. (2a) Band Diagram for AlAs.**

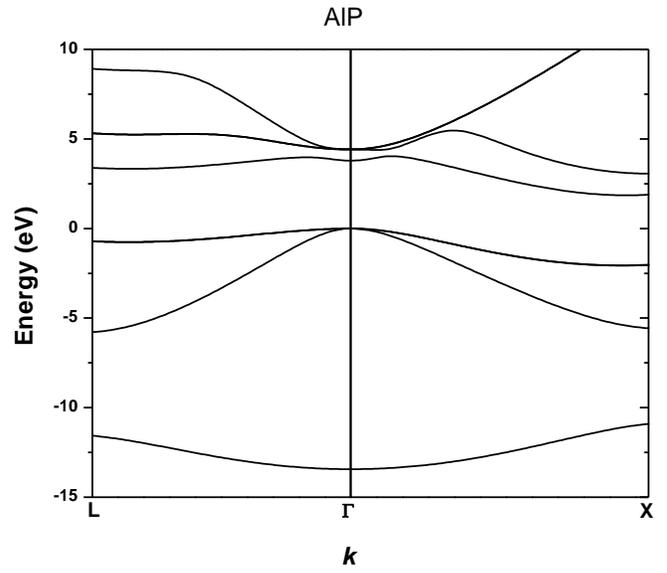

**Fig. (2b) Band Diagram for AlP.**

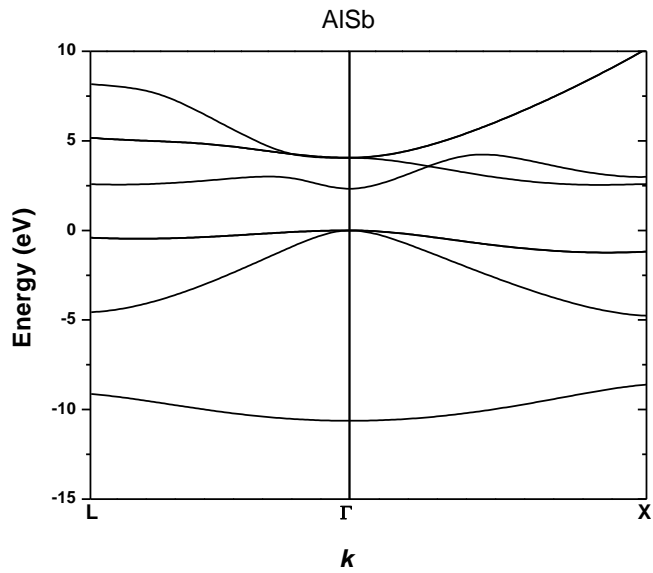

**Fig. (2c) Band Diagram for AlSb.**

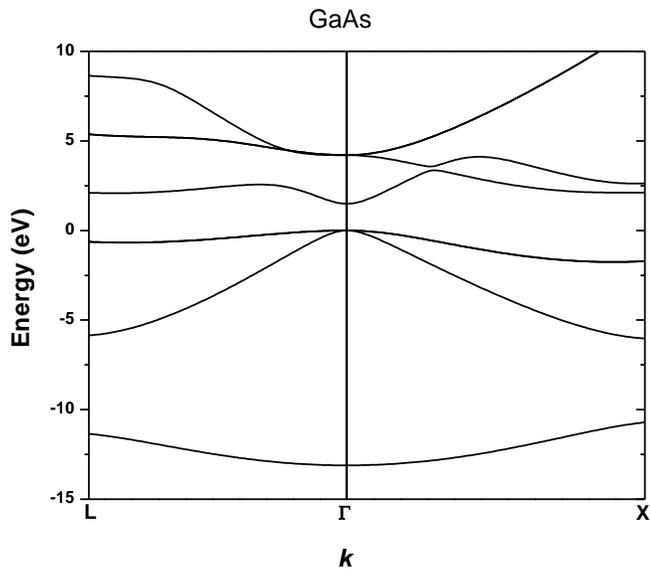
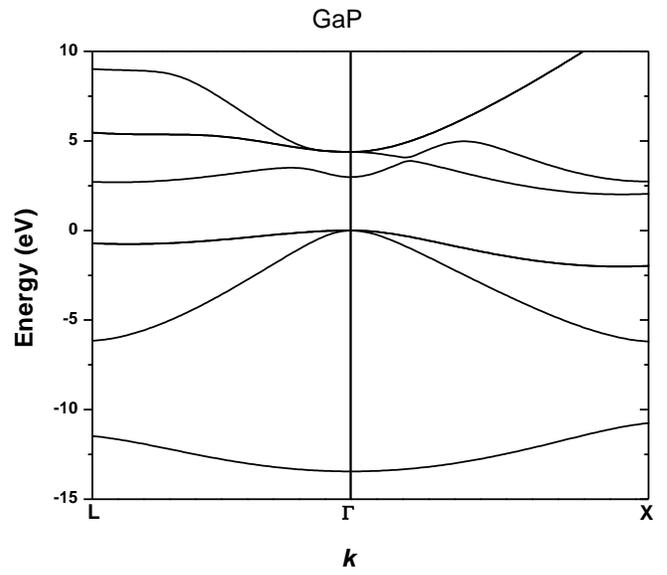

**Fig. (2d)   Band Diagram for GaAs.**                **Fig. (2e)  Band Diagram for GaP.**

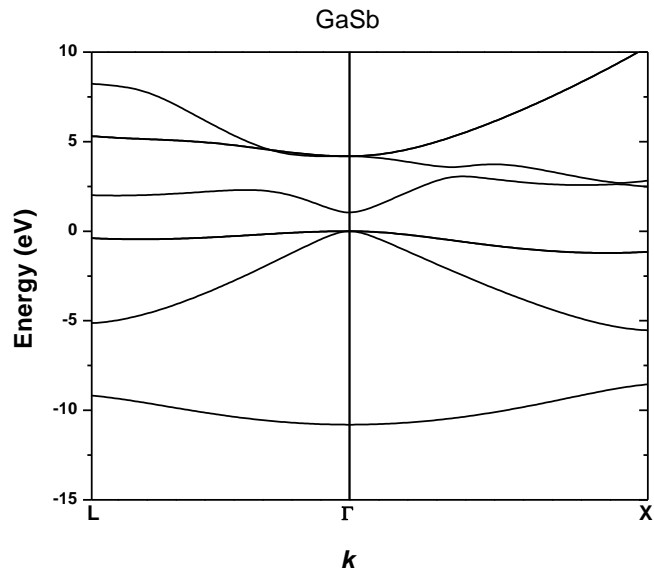

**Fig. (2f)  Band Diagram for GaSb.**

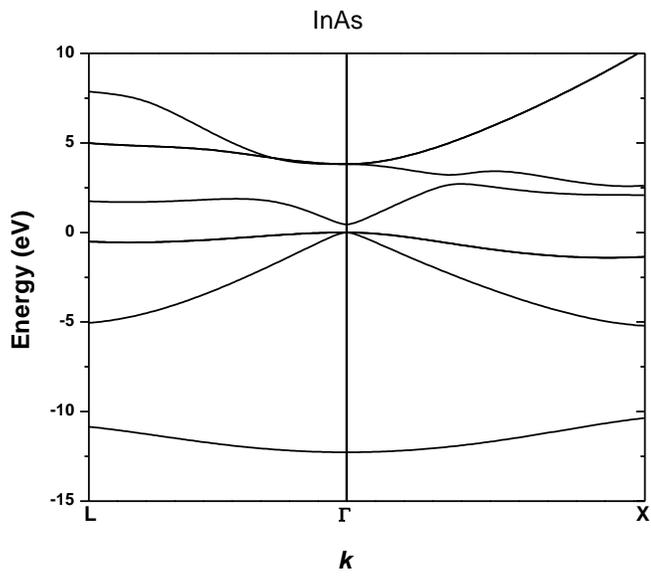

**Fig. (2g) Band Diagram for InAs.**

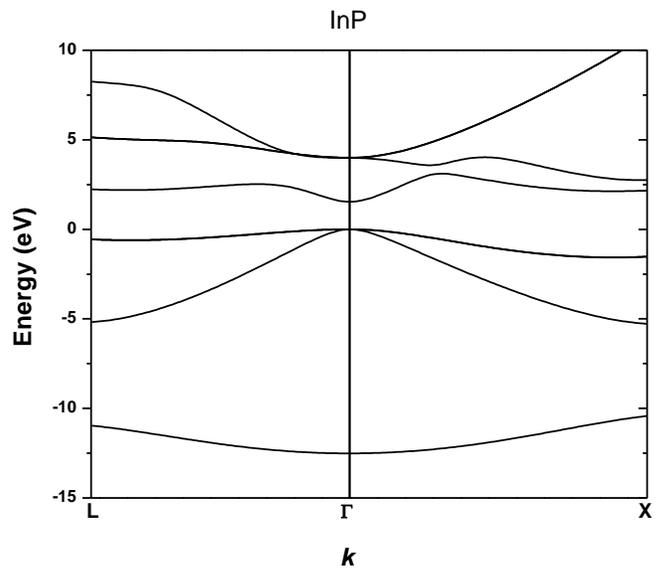

**Fig. (2h) Band Diagram for InP.**

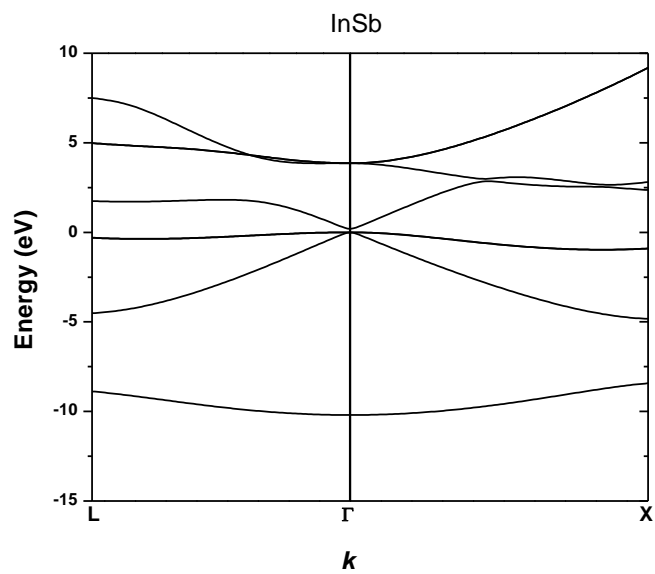

**Fig. (2i) Band Diagram for InSb.**

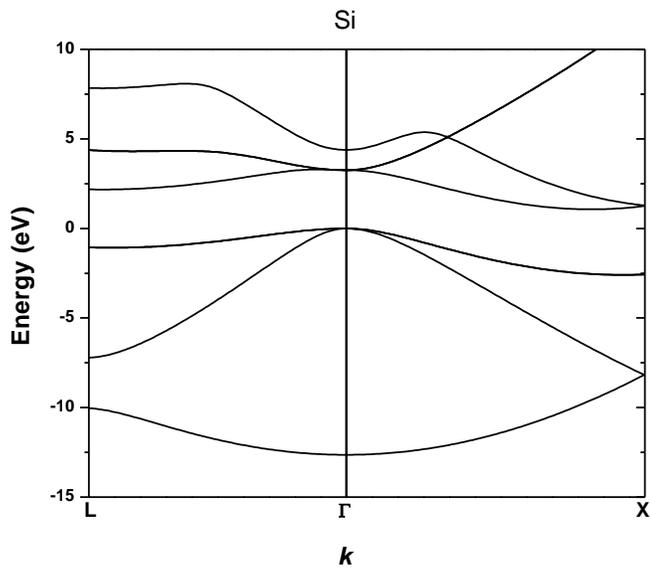

**Fig. (3a)   Band Diagram for Si.**

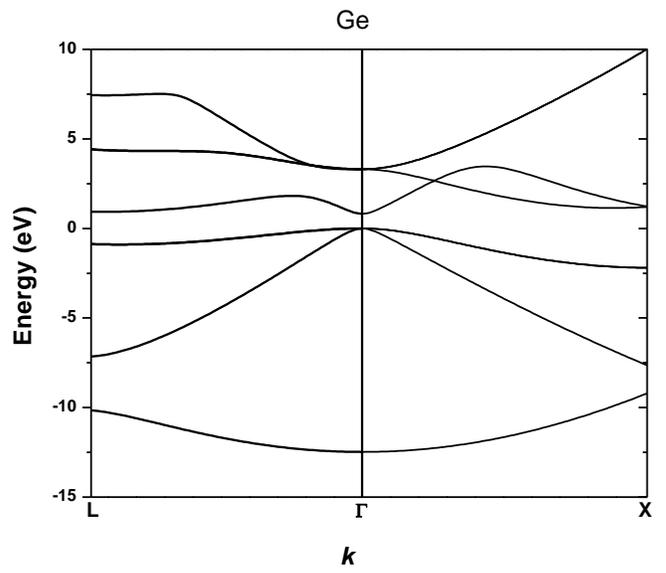

**Fig. (3b)  Band Diagram for Ge.**

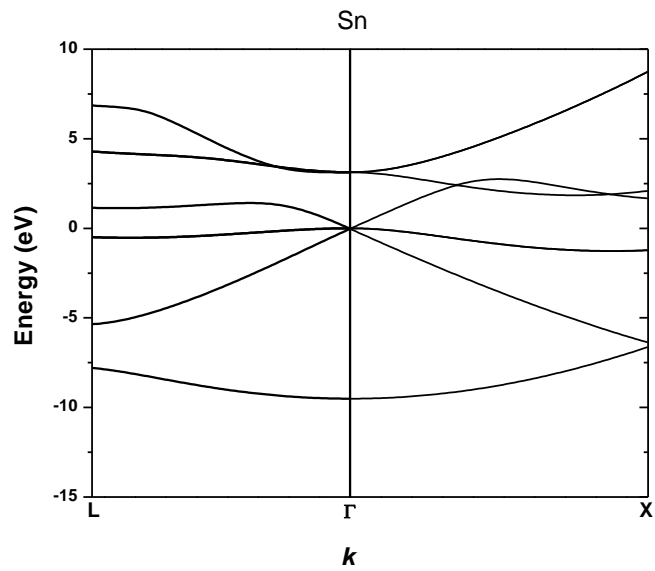

**Fig. (3c)  Band Diagram for Sn.**